\documentclass[11pt,twoside]{article}
\usepackage{amsmath}
\usepackage{amssymb}
\usepackage{amscd}
\usepackage{graphicx}
\usepackage{epstopdf}
\newcommand{\remark}{\smallbreak\noindent{\bf Remark.}}
\renewcommand{\a}{\alpha}
\renewcommand{\b}{\beta}
\newcommand{\g}{\gamma}
\newcommand{\G}{\Gamma}
\renewcommand{\d}{\delta}
\newcommand{\e}{\varepsilon}
\newcommand{\z}{{\zeta}}

\newcommand{\Th}{\Theta}
\newcommand{\io}{\iota}

\renewcommand{\l}{\lambda}
\newcommand{\m}{\mu}

\renewcommand{\r}{\rho}
\newcommand{\s}{\sigma}
\renewcommand{\t}{\tau}
\newcommand{\om}{{\omega}}
\newcommand{\Cs}{{\rlap{\lower3pt\hbox{\textnormal{\LARGE \char'040}}}{\Gamma}}{}}
\newcommand{\h}{\hbar}
\newcommand{\de}{\partial}
\newcommand{\GGs}{{\scriptscriptstyle{\mathbb{G}}}}
\newcommand{\oh}{\tfrac{1}{2}}
\newcommand{\oq}{\tfrac{1}{4}}
\newcommand{\osq}{\tfrac{1}{\surd2}}
\newcommand{\isq}{\tfrac{\iO}{\surd2}}

\newcommand{\thW}{\theta_{{\!}_{\text{\sc{w}}}}}
\newcommand{\cj}[1]{\overline{#1}}
\newcommand{\lin}{{\scriptscriptstyle\bigstar}}
\newcommand{\alin}{{\overline{\scriptscriptstyle\bigstar}}}
\renewcommand{\.}{{\scriptstyle\boldsymbol{\dot{}}}}
\newcommand{\td}{\tilde}
\newcommand{\cev}[1]{\overset{\smash{{}_{\gets}}}{#1}}
\newcommand{\Lll}{{\scriptscriptstyle{\mathrm{L}}}}
\newcommand{\Rrr}{{\scriptscriptstyle{\mathrm{R}}}}

\newcommand{\grav}{{}_{\mathrm{g}}}
\newcommand{\emag}{{}_{\mathrm{em}}}
\newcommand{\Dir}{{}_{\sst{\mathrm{D}}}}
\newcommand{\interaction}{{}_{\sst{\mathrm{int}}}}
\newcommand{\fl}{\flat}
\newcommand{\bl}{{\bar\lambda}}
\newcommand{\bh}{{\bar h}}
\newcommand{\bq}{{\bar q}}
\newcommand{\bs}{{\bar s}}
\newcommand{\bu}{{\bar u}}
\newcommand{\bv}{{\bar v}}
\newcommand{\bw}{{\bar w}}
\newcommand{\bz}{{\bar z}}
\newcommand{\bzz}{{\bar\zz}}
\newcommand{\be}{{\bar\varepsilon}}
\newcommand{\bze}{{\bar\zeta}}
\newcommand{\bch}{{\bar\chi}}
\newcommand{\bps}{{\bar\psi}}
\newcommand{\A}{{\boldsymbol{A}}}
\newcommand{\B}{{\boldsymbol{B}}}

\newcommand{\E}{{\boldsymbol{E}}}
\newcommand{\Ec}{\cj{\E}}
\newcommand{\F}{{\boldsymbol{F}}}
\newcommand{\Fl}{\F^\lin}
\newcommand{\Fc}{\cj{\F}}
\renewcommand{\H}{{\boldsymbol{H}}}
\newcommand{\I}{{\boldsymbol{I}}}

\newcommand{\Ic}{\cj{\I}}
\newcommand{\Ia}{\cj{\I}{}^\lin}
\newcommand{\Il}{\I{}^\lin}
\newcommand{\M}{{\boldsymbol{M}}}
\renewcommand{\P}{{\boldsymbol{P}}}
\newcommand{\Pm}{\P_{\!\!m}}
\newcommand{\Q}{{\boldsymbol{Q}}}
\newcommand{\Spin}{{\boldsymbol{S}}}
\newcommand{\Sc}{\cj{\Spin}}
\newcommand{\Sl}{\Spin{}^\lin}
\newcommand{\U}{{\boldsymbol{U}}}
\newcommand{\Uc}{\cj{\U}}

\newcommand{\Ua}{\cj{\U}{}^\lin}
\newcommand{\Ul}{\U{}^\lin}
\newcommand{\V}{{\boldsymbol{V}}}
\newcommand{\Vc}{\cj{\V}}
\newcommand{\Val}{\V{}^\alin}
\newcommand{\Va}{\cj{\V}{}^\lin}
\newcommand{\Vl}{\V{}^\lin}
\newcommand{\W}{{\boldsymbol{W}}}
\newcommand{\Wc}{\cj{\W}}
\newcommand{\Wa}{\cj{\W}{}^\lin}
\newcommand{\Wl}{\W{}^\lin}
\newcommand{\X}{{\boldsymbol{X}}}
\newcommand{\Z}{{\boldsymbol{Z}}}
\newcommand{\Zc}{\cj{\Z}}
\newcommand{\Ug}{\mathrm{U}}
\newcommand{\SU}{\mathrm{SU}}
\newcommand{\CC}{{\mathbb{C}}}
\newcommand{\II}{{\mathbb{I}}}
\newcommand{\LL}{{\mathbb{L}}}
\newcommand{\MM}{{\mathbb{M}}}
\newcommand{\NN}{{\mathbb{N}}}
\newcommand{\QQ}{{\mathbb{Q}}}
\newcommand{\RR}{{\mathbb{R}}}
\newcommand{\SSI}{{\mathbb{S}}}
\newcommand{\TT}{{\mathbb{T}}}
\newcommand{\UU}{{\mathbb{U}}}
\newcommand{\VV}{{\mathbb{V}}}
\newcommand{\ZZ}{{\mathbb{Z}}}
\newcommand{\Lcal}{{\mathcal{L}}}
\newcommand{\lfr}{\mathfrak{l}}
\newcommand{\Lfr}{\mathfrak{L}}
\newcommand{\End}{\operatorname{End}}
\newcommand{\Aut}{\operatorname{Aut}}

\newcommand{\Tr}{\operatorname{Tr}}
\newcommand{\Ker}{\operatorname{Ker}}
\newcommand{\ad}{\operatorname{ad}}
\newcommand{\Id}[1]{{1\!\!1}\!{}_{#1}{}}

\newcommand{\dO}{\mathrm{d}}
\newcommand{\KO}{\mathrm{K}}
\newcommand{\HO}{\mathrm{H}}
\newcommand{\kO}{\mathrm{k}}
\newcommand{\RO}{\mathrm{R}}
\newcommand{\TO}{\mathrm{T}}
\newcommand{\TS}{\TO^{*}\!}
\newcommand{\VO}{\mathrm{V}}

\newcommand{\dx}{\dO\xx}
\newcommand{\iO}{\mathrm{i}}
\newcommand{\na}{\nabla\!}
\newcommand{\nasl}{{\rlap{\raise1pt\hbox{\,/}}\nabla}}
\newcommand{\ten}[1]{\operatorname*{\otimes}_{\!{\scriptscriptstyle #1}} }
\newcommand{\cart}[1]{\operatorname*{\times}_{\!{\scriptscriptstyle #1}} }
\newcommand{\we}{{\,\wedge\,}}
\newcommand{\weu}[1]{{\wedge^{\!#1}}}
\newcommand{\pint}{\mathord{\rfloor}}
\newcommand{\comp}{\mathbin{\raisebox{1pt}{$\scriptstyle\circ$}}}
\newcommand{\tn}{{\,\otimes\,}}
\newcommand{\Bang}[1]{{\left\langle#1\right\rangle}}
\newcommand{\bang}[1]{{\langle#1\rangle}}
\newcommand{\Ii}[2]{{}^{#1}_{\phantom{#1}\!#2}}
\newcommand{\iI}[2]{{}_{#1}^{\phantom{#1}\!#2}}
\newcommand{\iIi}[3]{{}_{#1\phantom{#2}\!\!#3}^{\phantom{#1}\!#2}}
\newcommand{\si}[2]{\sigma\iI{#1}{#2}}
\newcommand{\sA}{{\scriptscriptstyle A}}
\newcommand{\sB}{{\scriptscriptstyle B}}
\newcommand{\cA}{{\sA\.}}
\newcommand{\cB}{{\sB\.}}
\newcommand{\AAd}{{\sA\cA}}
\newcommand{\BBd}{{\sB\cB}}
\newcommand{\zeA}{{\zeta_\sA}}
\newcommand{\bzeA}{{\bze_\cA}}

\newcommand{\zzA}{\zz^\sA}
\newcommand{\zzB}{\zz^\sB}

\newcommand{\bb}{{\mathsf{b}}}
\newcommand{\Ssf}{{\mathsf{S}}}
\renewcommand{\ss}{{\mathsf{s}}}

\newcommand{\xx}{{\mathsf{x}}}
\newcommand{\yy}{{\mathsf{y}}}
\newcommand{\zz}{{\mathsf{z}}}

\newcommand{\sst}{\scriptscriptstyle}
\newcommand{\sTh}{{\breve\Theta}}
\newcommand{\qRq}{{\quad\Rightarrow\quad}}
\newcommand{\into}{\hookrightarrow}
\newcommand{\onto}{\rightarrowtail}
\newcommand{\ul}{\underline}
\newcommand{\phvac}{\phi_0}
\title{Tetrad gravity, electroweak geometry and conformal symmetry}
\date{{\small v2 -- 2 November 2010} }
\author{Daniel Canarutto\\[6pt]
{\small\it Dipartimento di Matematica Applicata ``G. Sansone'', }\\
{\small\it Via S. Marta 3, 50139 Firenze, Italia}\\
{\small email:~daniel.canarutto@unifi.it}\\
{\small http://www.dma.unifi.it/\char126 canarutto}}
\begin{document}
\bibliographystyle{alpha}
\maketitle
\begin{abstract}\noindent
A partly original description of gauge fields
and electroweak geometry is proposed.
A discussion of the breaking of conformal symmetry
and the nature of the dilaton in the proposed setting
indicates that such questions cannot be definitely answered
in the context of electroweak geometry.
\end{abstract}



\tableofcontents

\vfill\newpage

\section*{Introduction}

Several papers discussing electroweak geometry,
and proposing modifications,
have appeared even recently~\cite{Fa,FKD,LM,ILM,MT,Mo10,RySh,Ta}.
Actually the geometric background of the EW theory,
in its current formulation,
lacks the tidiness of the Dirac theory of spinors.
One can see various unsatisfactory features:
the meaning of the mixing angle,
the origin of the potential term related to symmetry breaking,
the need for various \emph{ad hoc} choices.
Furthermore, from a physical point of view,
at least one crucial aspect of the electroweak theory
(the existence of the Higgs particle)
is still waiting for experimental confirmation,
and some parameters still have to be precisely measured.

In this paper I attempt to give a
partly original presentation of electroweak geometry,
in the hope that some features can be seen more clearly
by trying and somewhat changing the roles of various objects.

In \S\ref{s:Mathematical preliminaries}
I'll briefly review some basic mathematical notions which,
although by now widely discussed in the literature,
cannot be seen as standard knowledge;
in particular, \S\ref{ss:Unit spaces and scaling} contains a
``not too short'' discussion of the unit spaces approach to scaling.
In~\S\ref{s:Electrodynamics} I'll sketch the main ideas
of a certain two-spinor formulation of electrodynamics
in curved spacetime~\cite{C00b,C07}
which I regard as specially neat from a geometrical point of view.

Those ideas suggest a somewhat non-standard general approach,
discussed in~\S\ref{s:Gauge fields},
to gauge fields
and their relation to the classical geometry underlying a field theory.
Finally, in~\S\ref{s:Electroweak field theory},
I apply those ideas to a partly original formulation of electroweak geometry.
The different role of spin with respect to other internal
degrees of freedom is discussed,
and the Higgs field is seen to arise naturally in this context.

In \S\ref{ss:Electroweak geometry and the dilaton}
I discuss the question of the breaking of conformal invariance
and the nature of the dilaton in connection to
electroweak geometry and to the Higgs field in particular,
briefly commenting about some ideas which have appeared in the literature;
here I argue that the question cannot actually
be given a definite, convincing answer
in the context of electroweak geometry:
some substantial extension is needed.

By the way, it should be noted that the geometric language used here
is somewhat less group-oriented than standard approaches
(though eventually the two languages are essentially equivalent).

\section{Mathematical preliminaries}
\label{s:Mathematical preliminaries}

\subsection{Tangent-valued forms, brackets and connections}
\label{ss:Tangent-valued forms, brackets and connections}

We'll deal essentially with smooth finite-dimensional manifolds and bundles,
and smooth maps.
By $\TO$ and $\VO$ we denote the tangent and vertical functors.

The Fr\"olicher-Nijenhuis algebra of tangent-valued forms
provides us with a general framework for connections and related topics;
though based on firmly established results
in the literature~\cite{FN56,FN60,Mod91,MK,Mi01},
this framework may be less familiar to the reader
than the more usual language of principal bundles;
hence, a brief account of the basics could be appropriate.

A \emph{tangent valued $r$-form} on a manifold $\M$\,, $r\in\{0\}\cup\NN$\,,
is a (local) smooth section
$\M\to\weu{r}\TS\M\ten{\M}\TO\M$\,.
The sheaf of all tangent valued forms has a natural structure
of a graded Lie algebra determined by the \emph{Fr\"olicher-Nijenhuis bracket}.
If $\phi$ is a t.v.\ $r$-form and $\psi$ is a t.v.\ $s$-form
then their FN bracket $[\phi,\psi]$
is the t.v.\ $r{+}s$-form which has the coordinate expression\footnote{
The FN bracket can be characterized, in coordinate-free language, by
\begin{equation*}\begin{split}
[\a\tn u,\b\tn v]&=\a\we\b\tn[u,v]+\a\we(u.\b)\tn v-(v.\a)\we\b\tn u \\
  &\quad +(-1)^r(v|\a)\we\dO\b\tn u +(-1)^r\dO\a(u|\b)\tn v~,
\end{split}\end{equation*}
where $\a:\M\to\weu{r}\TS\M$, $\b:\M\to\weu{s}\TS\M$, $u,v:\M\to\TO\M$
and $[u,v]$ is the Lie bracket of $u$ and $v$\,.} 
\begin{multline*}
[\phi,\psi]_{a_1\dots a_{r+s}}^b=\bigl(
\phi_{a_1\dots a_r}^c\,\de_c\psi_{a_{r+1}\dots a_{r+s}}^b
-(-1)^{rs}\,\psi_{a_1\dots a_s}^c\,\de_c\phi_{a_{s+1}\dots a_{r+s}}^b
\\[6pt]
-r\,\phi_{a_1\dots a_{r-1}c}^b\,\de_{a_r}\psi_{a_{r+1}\dots a_{r+s}}^c
+(-1)^{rs}\,s\,\psi_{a_1\dots a_{s-1}c}^b\,
\de_{a_s}\phi_{a_{s+1}\dots a_{r+s}}^c \bigr)~.
\end{multline*}

The Fr\"olicher-Nijenhuis algebra on a fibered manifold $\E\onto\M$
has a special interest.
In general this is very complicate,
however one deals mostly with the subalgebra of \emph{projectable} forms.
A tangent valued $r$-form $\phi$ on $\E$ is said to be \emph{basic} if
it is a section
$$\phi:\E\to\weu{r}\TS\M\ten{\E}\TO\E\subset\weu{r}\TS\E\ten{\E}\TO\E~,$$
and \emph{projectable} over $\ul\phi:\M\to\weu{r}\TS\M\ten{\M}\TO\M$
if it is basic and the diagram
$$\begin{CD}
\E @>{\phi}>> \weu{r}\TS\M\ten{\E}\TO\E \\
@VVV @VVV \\
\M @>>{\ul\phi}> \weu{r}\TS\M\ten{\M}\TO\M
\end{CD}$$
commutes.
The FN bracket of two projectable tangent valued forms
turns out to be projectable.
In particular, vertical-valued basic forms $\E\to\weu{r}\TS\B\ten{\E}\VO\E$
are always projectable.

A \emph{connection} of a fibered manifold is a tangent $1$-form
which is projectable over the identity
$\M\to\TS\M\ten{\M}\TO\M$.
Its expression in fibered coordinates $(\xx^a,\yy^i)$ is of the type
$\g=\dx^a\tn(\de\xx_a+\g_a^i\,\de\yy_i)$\,,
with $\g\iI{a}{i}:\E\to\RR$\,.
Its \emph{curvature tensor} is defined to be the FN bracket
$$R[\g]:=-[\g,\g]:\E\to\weu2\TS\M\ten{\E}\VO\E~.$$

Let $\E\onto\M$ be a vector bundle.
A projectable t.v.\ form is said to be \emph{linear}
if it is a linear morphism over its projection $\ul\phi$\,.
The Fr\"olicher-Nijenhuis bracket of
linear projectable t.v.\ forms turns out to be linear.
In particular, a linear connection is characterized in linear fiber coordinates
by $\g_a^i=\g\iIi aij\,\yy^j$ with $\g\iIi aij:\M\to\RR$\,.
Its curvature tensor can be seen as a section
$$R[\g]:\E\to\weu2\TS\M\ten{\M}\End\E\cong
\weu2\TS\M\ten{\M}\E\ten{\M}\E^*~,$$
with coordinate expression
$R\iIi{ab}ij=-\de_a\g\iIi bij+\de_b\g\iIi aij
+\g\iIi aih\,\g\iIi bhj-\g\iIi bih\,\g\iIi ahj$\,.

For an arbitrary connection,
the \emph{covariant differential} of a section $s:\M\to\E$
is defined to be the section
$\nabla[\g]s:=\TO s-\g\comp s:\M\to\TS\M\ten{\E}\VO\E$\,,
with coordinate expression
$\na_a s^i=\de_a s^i-\g_a^i\comp s$\,.
In the linear case one has $\VO\E\cong\E\cart{\M}\E$,
so that $\nabla[\g]s$ can be seen as a section $\M\to\TS\M\ten{\M}\E$.

\subsection{Hermitian spaces}
\label{ss:Hermitian spaces}
If $\V$ is a complex vector space of finite-dimension $n$\,,
then we denote its \emph{dual} and \emph{antidual} spaces
respectively as $\Vl$ and $\V^\alin$,
mutually anti-isomorphic via the correspondence\footnote{
$\V^\alin$ is the space of all antilinear maps $\V\to\CC$\,.
In general, if $f$ is any function
then $\bar f:x\mapsto\overline{f(x)}$\,.} 
\hbox{$\l\mapsto\bl$}\,.
Moreover we indicate as $\Vc:=\V^{\lin\alin}$
the \emph{conjugate space} of $\V$.
Up to natural isomorphisms one gets only four distinct spaces
$\V\leftrightarrow\Vc$\,, $\Vl\leftrightarrow\Val\cong\Va$
(the arrows indicate the conjugation anti-isomorphisms).
Accordingly we use four different index types,\footnote{
Let $(\bb_\sA)$, $1\leq A\leq n$\,, be a basis of $\V$
and $(\bb^{\sA})$ its dual basis of $\Vl$.
Then we have the induced bases $\bigl(\bar\bb_{\cA}\bigr)$ of $\Vc$
and $\bigl(\bar\bb^{\cA}\bigr)$ of $\Va$,
with $\bar\bb_{\cA}:=\overline{\bb_\sA}$ and the like.
For $v\in\V$, for example, we have
\hbox{$v=v^\sA\,\bb_\sA$} and \hbox{$\bv=\bv^\cA\,\bar\bb_\cA$}\,,
with $\bv^\cA=\overline{v^\sA}$\,.
The conjugation morphism can be extended to tensors of any rank and type,
exchanging dotted and non-dotted index types.
Observe that dotted indices cannot be contracted with non-dotted indices.
In particular if $K\in\Aut(\V)\subset\V\tn\Vl$
then $\bar K\in\Aut(\Vc)\subset\Vc\tn\Va$ is the induced
conjugated transformation
(under a basis transformation,
dotted indices transform with the conjugate matrix).} 
with `dotted' indices referring to `conjugated' spaces $\Vc$ and $\Va$.

The space $\V\tn\Vc$ has a natural real linear (complex anti-linear) involution
$w\mapsto w^\dag$,
which on decomposable tensors reads
\hbox{$(u\tn\bv)^\dag:=v\tn\bu$}\,.
Hence one has the natural decomposition of $\V\tn\Vc$ into the direct sum
of the \emph{real} eigenspaces of the involution with eigenvalues $\pm1$,
respectively called the \emph{Hermitian} and \emph{anti-Hermitian} subspaces,
namely
$$\V\tn\Vc=\HO(\V\tn\Vc)\oplus \iO\,\HO(\V\tn\Vc)~.$$
In other terms, the Hermitian subspace $\HO(\V\tn\Vc)$ is constituted by
all $w\in\V\tn\Vc$ such that $w^\dag=w$,
while an arbitrary $w$ is uniquely decomposed into the sum of an Hermitian
and an anti-Hermitian tensor as
$$w=\oh(w+w^\dag)+\oh(w-w^\dag)~.$$
Then $w=w^{\sA\cB}\bb_\sA\tn\bar\bb_\cB$ is Hermitian (anti-Hermitian)
iff the matrix $(w^{\sA\cB}\,)$ of its components is such,
namely $\bw^{\cB\sA}=\pm w^{\sA\cB}$.
One has the natural isomorphism
$[\HO(\V\tn\Vc)]^*\cong\HO(\Vl\tn\Va)$\,,
where ${}^*$ denotes the \emph{real} dual.

A Hermitian $2$-form is defined to be a Hermitian tensor $h\in\HO(\Va\tn\Vl)$.
The associated quadratic form $v\mapsto h(v,v)$ is real-valued.
The notions of signature and non-degeneracy of Hermitian $2$-forms
are introduced similarly to the case of real bilinear forms.
If $h$ is non-degenerate then it yields the isomorphisms
$h^\fl:\Vc\to\Vl:\bv\mapsto h(\bv,\_)$
and
$\bh^\fl:\V\to\Va:v\mapsto h(\_,v)$.
The inverse isomorphisms are denoted as $h^\#$ and $\bh^\#$\,.

\subsection{Unit spaces and scaling}
\label{ss:Unit spaces and scaling}

An algebraically precise treatment of physical scales
was introduced around 1995 after an idea of M.\:Modugno,
and has been systematically used, since then,
in papers of various authors~\cite{CJM95,JM02,JM06,JMV10,MST05,SV00,Vi99,Vi00}.
The basic notion is that of a \emph{positive space}
(or \emph{scale space}, or \emph{unit space}),
namely a 1-dimensional `semi-vector space' without the zero element.
Though physical scales (or `dimensions') are usually dealt with
in an `informal' way, without a precise mathematical setting,
the notion of a scale space arises naturally from simple arguments.
The distance of two points in Euclidean space, for example,
can be expressed as a real number
only if a length unit has been fixed,
and the set $\LL$ of lengths is naturally endowed
with a free and transitive left action $\RR^+\times\LL\to\LL$\,;
this determines an algebraic structure of semi-vector space over $\RR^+$
(note that $\LL$ has no distinguished element).

A rigorous study of this matter~\cite{JMV10}
turns out to be more delicate than one may expect at first sight,
but the basic notions needed for ``everyday use'' can be easily sketched.

A \emph{semi-vector space} is defined to be a set $\A$
equipped with an addition map $\A\times\A\to\A$
and a multiplication map $\RR^+\times\A\to\A$\,,
fulfilling the usual axioms of vector spaces
except those properties which involve opposites
and the zero element.
Then, in particular, any vector space is a semi-vector space,
and the set of linear combinations over $\RR^+$
of $n$ fixed independent vectors
in a vector space is a semi-vector space.
\emph{Semi-linear} maps between semi-vector spaces
are defined in an obvious way;
in particular we obtain the \emph{semi-dual space} $\A^*$
(or simply the `dual space') of any semi-vector space.

A semi-vector space $\UU$ is called a \emph{positive space}
if the multiplication $\RR^+\times\UU\to\UU$ is a transitive
left action of the multiplicative group $\RR^+$ on $\UU$
(then a positive space cannot have a zero element).
If $b\in\UU$ then any other element $u\in\UU$
can be written as $u^0\,b$ with $u^0\in\RR^+$.
Quite naturally we can write $u^0\equiv u/b$\,, that is
$u=(u/b)\,b$\,, $(u/b)\in\RR^+$\,.
So we might also say that a positive space
is a `1-dimensional' semi-vector space.

Several concepts and results of standard linear and multi-linear algebra
related to vector spaces can be easily repeated for semi-vector spaces
and positive semi-vector spaces
(including linear and multi-linear maps, bases, dimension, 
tensor products and duality).
The main caution to be taken is to avoid formulations
which involve the zero element.
In particular,
one can define the tensor product (over $\RR^+$) of semi-vector spaces;
the tensor product of a semi-vector space and a real or complex vector space
becomes naturally also a vector space.

A $1$-dimensional semi-vector space will be called a \emph{unit space}.
In particular, let $\UU$ be a positive unit space;
then the trace yields the natural identification $\UU\tn\UU^{*}\cong\RR^+$.
Moreover for any $m\in\NN$ there is, up to isomorphism,
a unique \emph{$m$-root} of $\UU$;
this is a positive unit space, denoted as $\UU^{1/m}$,
with the property ${\otimes}^m(\UU^{1/m})\cong\UU$.
Introducing the notation
\begin{align*}&
\UU^{p/m}:={\otimes}^p(\UU^{1/m})\cong({\otimes}^p\UU)^{1/m}~,\quad p\in\NN~,
\\&
\UU^{-1}:=\UU^{*}~,
\end{align*}
one obtains the definition of $\UU^r$ for all rational exponents $r$.

Accordingly, one may adopt a `number-like' notation for elements of unit spaces.
Namely one writes $u^{-1}:=u^*\in\UU^{-1}$ (the dual element)
and $uv:=u\tn v$ for $u\in\UU$, $v\in\VV$.
This allows a rigorous treatment of measure units in physics,
maintaining a notation close to the traditional one.

In many physical theories it is convenient to assume
the spaces $\TT$ of \emph{time scales},
$\LL$ of \emph{length scales} and $\MM$ of \emph{mass scales}
as the \emph{basic spaces of scales}.
An arbitrary \emph{scale space} is defined to be
a positive space of the type
$\SSI=\TT^{d_1}\tn\LL^{d_2}\tn\MM^{d_3}$\,, with $d_i\in\QQ$\,.
An element $\ss\in\SSI$ is also called a \emph{scale},
or a \emph{unit of measurement}.

A `scaled' version of a vector bundle $\E\to\B$
is a fibered tensor product $\SSI\tn\E\to\B$\,.
Fibered linear algebraic or differential
operations on $\E$ determine analogue scaled operations.
In particular, a linear connection  $\G$ of $\E\to\B$
determines a linear connection of $\SSI\tn\E\to\B$\,.

Two sections $\s:\B\to\E$ and $\s':\B\to\SSI\tn\E$
of differently scaled vector bundles
can be compared if we avail of a scale factor $\ss:\B\to\SSI$\,,
called a \emph{coupling scale},
or possibly a  \emph{coupling constant}.
The commonest coupling constants are
the {speed of light} $c\in\TT^{-1}\tn\LL$\,;
{Planck's constant} $\h\in\TT^{-1}\tn\LL^2\tn\MM$\,;
Newton's {gravitational constant}
${\scriptstyle{\mathbb{G}}}\in\TT^{-2}\tn\LL^3\tn\MM^{-1}$\,;
the {positron charge} $e\in\TT^{-1}\tn\LL^{3/2}\tn\MM^{1/2}$\,;
a {particle's mass} $m\in\MM$\,.
By viewing $c$ as an isomorphism $\TT\to\LL$
and then $\h$ as an isomorphism $\MM\to\LL^*$
one can express all physical scales as powers of $\LL$
(this is the familiar `natural units' setting,
usually introduced by the condition $c=\h=1$).

Certain scale spaces arise quite naturally.
If $\X$ is an $n$-dimensional real vector space
then the choice of an orientation amounts to the choice
of a positive subspace of $\weu{n}\X$.
In the context of complex geometry we have another interesting construction:
if $\Z$ is a $1$-dimensional complex vector space
then the Hermitian subspace $\HO(\Z\tn\Zc)\subset\Z\tn\Zc$
(\S\ref{ss:Hermitian spaces}) is a 1-dimensional \emph{real} vector space 
which has the \emph{distinguished} positive subspace
$$\HO(\Z\tn\Zc)^+:=\bigl\{z\tn\bz:z\in\Z\setminus\{0\}\bigr\}~.$$
Thus, any $n$-dimensional complex vector space $\V$ yields
the positive space $\HO(\weu{n}\V\tn\weu{n}\Vc)^+$.

The metric, either Euclidean or Lorentzian,
is appropriately described as a scaled tensor field.
Focusing our attention on the spacetime $(\M,g)$ of General Relativity,
the metric is a section
$$g:\M\to\LL^2\tn\TS\M\tn\TS\M~,$$
so that the scalar product of vectors is valued into
$\RR\tn\LL^2\equiv\RR\tn\LL\tn\LL$\,.
Correspondingly, the volume form induced by $g$ is a scaled 4-form
$\eta:\M\to\LL^4\tn\weu4(\TS\M)$\,.
Note that $g$ and $\eta$ can also be seen as unscaled objects
on the fibers of
$\H\equiv\LL^{-1}\tn\TO\M\onto\M$.

The notion of a positive space can be extended
to that of a positive bundle over $\M$.
This will be the natural context in which one may address the questions
of running constants, conformal symmetry and the like
(\S\ref{ss:Electroweak geometry and the dilaton}).
By the way, we observe that the assignment of an orientation of $\M$
amounts to the choice of a positive bundle
$$\VV\M\equiv(\weu4\TO\M)^+\subset\weu4\TO\M~,$$
so that $\eta:\M\to\LL^4\tn\VV^{-1}\M$.
A section $\M\to\RR\tn\VV^{-1/2}\M$ is called a \emph{half-density}.

\section{Electrodynamics}
\label{s:Electrodynamics}
\subsection{Two-spinors and Lorentzian geometry}
\label{ss:Two-spinors and Lorentzian geometry}

The geometry of Dirac fields can be conveniently expressed
in the language of $2$-spinors.
In previous papers \cite{C98,C00b,C07} I treated such matters
according a partly original approach which uses minimal geometric assumptions;
this approach has a number of differences
with the classical Penrose formalism~\cite{PR84,PR88},
and includes an integrated treatment of
Einstein-Cartan-Maxwell-Dirac fields (see also~\cite{HCMN}).
The starting point is the realization that
a $2$-dimensional complex vector space $\Spin$,
with no further assumptions,
automatically generates a rich algebraic structure.
We begin with a brief account of that.
\smallbreak\noindent
$\bullet~$%
The Hermitian subspace of \hbox{$\weu2\Spin\tn\weu2\Sc$}
is a real $1$-dimensional vector space with a distinguished orientation;
its positively oriented semispace $\LL^2$
(whose elements are of the type $w\tn\bw$\,, $w\in\weu2\Spin$)
has the square root semispace $\LL$,
which will can be identified with the space of \emph{length units}.
\smallbreak\noindent
$\bullet~$%
The $2$-spinor space is defined to be $\U:=\LL^{-1/2}\tn\Spin$.
The space $\weu2\U$ is then naturally endowed with a Hermitian metric,
defined as the identity element in
$$\HO[(\weu{2}\Ua)\tn(\weu{2}\Ul)]
\cong\LL^2\tn\HO[(\weu{2}\Sl)\tn(\weu{2}\Sl)]~,$$
so that normalized `symplectic forms' $\e\in\weu2\Ul$
constitute a $\Ug(1)$-space (any two of them are related by a phase factor).
Each $\e$ yields the isomorphism
\hbox{$\e^\fl:\U\to\Ul:u\mapsto u^\fl:=\e(u,\_)$}\,.
\smallbreak\noindent
$\bullet~$%
The identity element in $\HO[(\weu{2}\Ua)\tn(\weu{2}\Ul)]$
can be written as $\e\tn\be$ where $\e\in\weu2\Ul$
is any normalized element.
This natural object can also be seen as a bilinear form $g$ on $\U\tn\Uc$,
via the rule
\hbox{$g(p\tn\bq,r\tn\bs)=\e(p,r)\,\be(\bq,\bs)$}
extended by linearity.
Its restriction to the Hermitian subspace $\H\equiv\HO(\U\tn\Uc)$
turns out to be a Lorentz metric.
Null elements in $\H$ are of the form $\pm u\tn\bu$ with $u\in\U$
(thus there is a distinguished time-orientation in $\H$).
\smallbreak\noindent
$\bullet~$%
Let $\W\equiv\U\oplus\Ua$.
The linear map
$\g:\U\tn\Uc\to\End(\W):y\mapsto\g(y)$ acting as
$$\td\g(p\tn\bq)(u,\chi)
=\sqrt2\bigl(\bang{\chi,\bq}\,p\,,\bang{p^\fl,u}\,\bq^\fl\,\bigr)$$
is well-defined independently of the choice of the normalized $\e\in\weu2\Ul$
yielding the isomorphism $\e^\fl$.
Its restriction to $\H$ turns out to be a Clifford map.
Thus one is led to regard $\W\equiv\U\oplus\Ua$ as the space of Dirac spinors,
decomposed into its Weyl subspaces.
The anti-isomorphism $\W\to\Wl:(u,\chi)\mapsto(\bch,\bu)$
is called the \emph{Dirac adjunction}
(\hbox{$\psi\mapsto\bar\psi$} in traditional notation),
and is associated with a natural Hermitian structure $\kO\in\Wa\tn\Wl$
which turns out to have the signature $(++--)$.

\medbreak
The above constructions and results can be read in coordinates as follows.
Consider an arbitrary basis $(\xi_\sA)$ of $\Spin$, $\scriptstyle{A}=1,2$\,.
This yields induced bases of the various associated spaces.
In particular we consider the bases $l\in\LL$ (a length unit),
$\bigl(\zeA\bigr)\equiv\bigl(l^{-1/2}\,\xi_\sA \bigr)\subset\U$,
$\e\in\weu2\Ul$.
We have $\e=\e_{\sA\sB}\,\zzA\we\zzB$\,,
where $\bigl(\zzA\bigr)\subset\Ul$ is the dual basis of $\bigl(\zeA\bigr)$
and $\bigl(\e_{\sA\sB}\bigr)$
denotes the antisymmetric Ricci matrix.\footnote{
In contrast to the usual $2$-spinor formalism,
no symplectic form is fixed.
The  $2$-form $\e$ is unique up to a phase factor
which depends on the chosen 2-spinor basis,
and determines isomorphisms
\begin{align*}
& \e^\fl:\U\to\Ul:u\mapsto u^\fl~,~~
\bang{u^\fl,v}:=\e(u,v)\qRq (u^\fl)_\sB=\e_{\sA\sB}\,v^\sA~,\\
& \e^\#:\Ul\to\U:\l\mapsto\l^\#~,~~
\bang{\m,\l^\#}:=\e^{-1}(\l,\m)\qRq (\l^\#)^\sB=\e^{\sA\sB}\,\l_\sA~.
\end{align*}} 
For $w\in\U\tn\Uc$ we write $w=w^{\AAd}\,\zeA\tn\bzeA$,
and get
$g(w,w)=\e_{\sA\sB}\,\be_{\cA\cB}\,w^{\AAd}\,w^{\BBd}=2\,\det w$\,.

As for the basis of $\H\equiv\HO(\U\tn\Uc)$ associated with $(\zeA)$
one usually considers the \emph{Pauli basis} $\bigl(\t_\l\bigr)$\,,
given by
$\t_\l\equiv\osq\,\si{\l}{\AAd}\,\zeA\tn\bzeA$
where $(\si{\l}{\AAd})$\,, $\l=0,1,2,3$\,, denotes the $\l$-th Pauli matrix.
This is readily seen to be orthonormal, namely
$g_{\l\m}\equiv g(\t_\l\,,\t_\m)=2\,\d^0_\l\d^0_\m-\d_{\l\m}$\,.

The associated \emph{Weyl basis} of $\W$
is defined to be the basis $(\z_\a)$, $\a=1,2,3,4$, given by
$$(\z_1\,,\z_2\,,\z_3,\z_4):=(\z_1\,,\z_2\,,-\bzz^1,-\bzz^2)~.$$
Above, $\z_1$ is a simplified notation for $(\z_1\,,0)$, and the like.
Another important basis is the \emph{Dirac basis} $(\z'_\a)$, $\a=1,2,3,4$,
where
$$\z'_1\equiv\osq(\z_1-\z_3)~,~~\z'_2\equiv\osq\,(\z_2-\z_4)~,~~
\z'_3\equiv\osq(\z_1+\z_3)~,~~\z'_4\equiv\osq(\z_2+\z_4)~.$$
Setting
$$\g_\l:=\g(\t_\l)\in\End(\W)$$
one recovers the usual Weyl and Dirac representations as the matrices
$\bigl(\g_\l\bigr)$\,, $\l=0,1,2,3$\,,
in the Weyl and Dirac bases respectively.

\medbreak
In standard expositions of electrodynamics one usually considers
some other structures which, however,
should not be seen as part of the basic assumptions
but rather depend on further choices.
In particular the assignment of a positive Hermitian metric on $\U$
and consequently on $\W$
(the related adjunction map is usually denoted as $\psi\mapsto\psi^\dagger$)
is readily seen to be equivalent to the assignment of an \emph{observer},
that is of a timelike future-oriented element in $\H^*\equiv\HO(\Ua\tn\Ul)$\,.
Also \emph{parity} is associated with the choice of an observer.
\emph{Charge conjugation}, on the other hand,
is the anti-involution $(u,\chi)\mapsto(\e^\#\bch,\be^\fl\bu)$
associated with the choice of a normalized 2-form $\e$\,,
and as such is unique up to a phase factor.
\emph{Time reversal} requires both
an observer \emph{and} a normalized symplectic form.

\subsection{Two-spinors and Einstein-Cartan-Maxwell-Dirac fields}
\label{ss:Two-spinors and Einstein-Cartan-Maxwell-Dirac fields}

Let $\Spin\onto\M$ be a complex vector bundle with $2$-dimensional fibers,
and consider the vector bundles obtained by performing
the above constructions fiberwise.
A linear connection $\Cs$ on $\Spin$ determines linear connections
on the associated bundles, and, in particular, connections
$G$ of $\LL$, $Y$ of $\weu2\U$ and $\td\G$ of $\H$.
Conversely $\Cs$ can be expressed in terms of these as
$$\Cs\iIi{a}{\sA}{\sB}=(G_a+\iO\,Y_a)\d\Ii{\sA}{\sB}
+\oh\,\td\G\iIi{a}{\sA\cA}{\sB\cA}
\equiv(G_a+\iO\,Y_a)\d\Ii{\sA}{\sB}+\td\Cs\iIi a\sA\sB~,$$
with $G_a\,,Y_a:\M\to\RR$\,, $\td\Cs\iIi a\sA\sB:\M\to\CC$\,,
$\td\Cs\iIi a\sA\sA=0$\,.

If $\M$ is $4$-dimensional then
a \emph{tetrad} (or \emph{soldering form})
is defined to be a linear morphism $\Th:\TO\M\to\LL\tn\H$.
An invertible tetrad determines, by pull-back,
a scaled Lorentz metric $\Th^*g$ on $\M$
and a metric connection of $\TO\M\onto\M$,
as well as a scaled Dirac morphism $\g\comp\Th:\TO\M\to\LL\tn\End\W$.
The \emph{Dirac operator}
$\psi\mapsto\nasl\psi$
is defined by natural contractions in $\cev\Th\tn(\g\nabla\psi)$\,,
where $\cev\Th:\LL\tn\H\to\TO\M$ is the inverse morphism of $\Th$\,.

In the above geometric environment we can formulate a non-singular field theory
even if $\Th$ is not required to be invertible everywhere~\cite{C98}.
If the invertibility requirement is satisfied
then the theory turns out to be essentially equivalent
to the standard theory of Einstein-Cartan-Maxwell-Dirac fields,
with some redefinition of the fundamental fields:
these are now the $2$-spinor connection $\Cs$,
the tetrad $\Th$,
the electromagnetic field $F:\M\to\LL^{-2}\tn\weu2\H^*$
and the Dirac field $\psi:\M\to\LL^{-3/2}\tn\W$.
The gravitational field is represented by the couple $(\Th,\td\G)$\,.
The connection $G$ induced on $\LL\onto\M$
is assumed to have vanishing curvature,
$\dO G=0$, so that we can find local charts such that $G_a=0$\,;
this amounts to `gauging away' the conformal `dilaton' symmetry.
Coupling constants then arise as covariantly constants sections of $\LL^r$,
$r\in\QQ$\,.

A natural unscaled Lagrangian density
depending on the said fields can be now introduced.
This uses the following observation:
if $\xi:\M\to\weu{r}\TS\M\tn\weu{r}\H$, $r=1,2,3$\,, then
$$\Th^{(4-r)}\we\xi:\M\to\LL^{4-r}\tn\weu{4}\TS\M\tn\weu{4}\H~,
\qquad \Th^{(p)}\equiv\Th\we\cdots\we\Th~~(p~\text{factors});$$
by contraction with the volume form of the fibers of $\H$
(determined by the Lorentz structure)
we then get a scaled density
$\sTh(\xi):\M\to\LL^{4-r}\tn\weu{4}\TS\M$.
So we obtain a Lagrangian density
$\Lcal=\Lcal\grav+\Lcal\emag+\Lcal\Dir$\,, with
\begin{align*}&
\Lcal\grav=\tfrac1{\GGs}\,\sTh(R[\td\G])~,\quad
\Lcal\emag=\bigl(\oq\,F^2-\oh\,\sTh(\dO Y\tn F)\bigr)\,\eta~,\quad
\\[6pt]&
\Lcal\Dir=\isq\,\sTh(\bang{\bps,\nabla\psi}-\bang{\nabla\bps,\psi})
-m\,\bang{\bps,\psi}\,\eta~.
\end{align*}
The details of the above expressions and of the calculations
of the Euler-Lagrange operator~\cite{C98,C00b}
are not essential here.
Eventually one gets the following field equations:
the Einstein equation and the equation for torsion;
the equation $F=2\,\dO Y$
(thus $Y$ is identified with the electromagnetic potential
up to a charge factor)
and the other Maxwell equation;
the Dirac equation,
also involving the torsion of the induced spacetime connection.

\subsection{Tetrad and QED interactions}\label{ss:Tetrad and QED interactions}

Generally speaking,
perturbative QFT requires certain basic ingredients:
time (namely the choice of some kind of observer),
\emph{free-particle states} and the \emph{quantum interaction};
the latter, on turn, is formed out of a specifically quantum ingredient
(a certain distribution on the bundle of particle momenta)
and of the classical interaction deduced from the Lagrangian
of the classical field theory
(essentially, a distinguished feature
of the underlying finite-dimensional geometric structure).
We are not going into many details here;
rather we'll make a few observations,
about the classical part of the interaction,
which will be relevant in the subsequent discussion.

Let $(M,g)$ be a Lorentzian spacetime
and $\Pm\subset\TS\M$ the subbundle over $\M$
whose fibers are the future hyperboloids (`mass-shells')
corresponding to mass $m\in\{0\}\cup\LL^{-1}$.
If $p\in(\Pm)_x$\,, $x\in\M$, then
$$\W\!\!_x=\W^{+}_{\!\!p}\oplus\W^{-}_{\!\!p}~,\quad
\W^{\pm}_{\!\!p}:=\Ker(\g[p^\#]\mp m)~,$$
where $p^\#\equiv g^\#(p)\in\LL^{-2}\tn\TS\M$
is the contravariant form of $p$\,.
Thus one has 2-fibered bundles $\W_{\!\!m}^\pm\to\Pm\to\M$, where
$$\W_{\!\!m}^\pm:=\bigsqcup_{p\in\Pm}\W_{\!\!p}^\pm\subset\Pm\cart{\M}\W~.$$
We call $\W^{+}_{\!\!m}$ and $\Wc{}^{-}_{\!\!m}$
the \emph{electron bundle} and the \emph{positron bundle},
respectively.\footnote{
Electron and positron quantum states can be then respectively introduced
as certain $\W^{+}_{\!\!m}$-valued and $\Wc{}^{-}_{\!\!m}$-valued
distributions on $\Pm$\,.
Details can be seen in a previous paper~\cite{C05}
where I presented some basic ideas
regarding quantum interactions, and QED in particular.} 

In standard presentations the classical interaction among
electron, positron and electromagnetic field
is read from the Dirac Lagrangian as the term
$-e\,\bang{\bps,\g[A]\psi}\equiv-\kO(\bps,\g[A]\psi)$\,,
where $e$ is the positron's charge
($e\,A=Y$, \S\ref{ss:Two-spinors and Einstein-Cartan-Maxwell-Dirac fields}).
When an electromagnetic gauge is chosen (see also~\S\ref{s:Gauge fields})
then $A$ can be viewed as a spacetime 1-form and,
through the tetrad and the metric,
as a section $\M\to\LL^{-1}\tn\H$.
Then we see that the classical interaction
is related to a natural contraction
$\ell\interaction:\Wc\cart{\M}\H\cart{\M}\W\to\CC$\,, that is a tensor field
$$\ell\interaction:\M\to\Wa\ten{\M}\H^*\ten{\M}\Wl~,\quad
\bang{\ell\interaction\,,\bar\phi\tn A\tn\psi}=-\kO(\bar\phi,\g[A]\psi)~.$$
By using the Hermitian structure $\kO$ of $\W$
(\S\ref{ss:Two-spinors and Lorentzian geometry})
and the metric $g$ of $\H$
one can ``raise indices'' in $\ell\interaction$\,,
thus obtaining eight ``clones'' of this tensor, with different index types.
In each type, a contravariant index corresponds to the creation of a particle,
and a covariant index corresponds to annihilation.
This mechanism
(together with its quantum counterpart acting on particle momenta)
essentially determines the allowed particle interactions.

\remark~
A peculiar aspect of the resulting quantum theory is that
the interactions can be written
solely in terms of ``internal'' spinor variables:
up to rearrangements of the index positions
we may only have contractions of the types
$u^\sB A_\BBd\,\bu^\cB$ and $\bch_\sB A^\BBd\chi_\cB$\,.
Furthermore, also the momentum variables in the propagators
can be expressed in terms of spinor variables.
In other words, any Feynman diagram can be seen as ``living''
just in the space of the ``internal'' particle states,
with the soldering form $\Th$ (the ``tetrad'')
eventually relating it to spacetime geometry.

\section{Gauge fields}
\label{s:Gauge fields}
\subsection{Gauge fields in classical and pre-quantum theories}
\label{ss:Gauge fields in classical and pre-quantum theories}

In a classical theory,
``matter fields'' and ``gauge fields'' are respectively described
as a section and a connection of some bundle $\E\onto\M$
(where $\M$ is the spacetime manifold).
If one assumes that the true physical meaning of the gauge fields
is encoded in the curvature tensor,
then one sees the connection itself as partly undetermined,
in the sense that different connections can yield the same physical field.
This is the essence of ``gauge freedom''.

In the corresponding theory of quantum particles
the situation is different.
In order to treat quantum interactions
one has to ``choose a gauge'',
so that the gauge field becomes a section of a vector bundle
and gets more degrees of freedom;
for real (asymptotic) gauge particles these
must be reduced by a suitable formalism.

In this paper I won't discuss constraints,
BRST symmetry or similar developments;
rather I'll take the provisional attitude
of considering the classical description of gauge fields as connections,
and the pre-quantum description of gauge fields in terms of
unconstrained sections of certain vector bundles,
as \emph{complementary} descriptions.

In standard presentations,
the choice of a gauge essentially amounts to the choice
of a local ``flat'' connection $\g_0$
(vanishing curvature tensor).
Then an arbitrary connection $\g$ is characterized by the difference
$\a\equiv\g\,{-}\,\g_0$\,, a true tensor field.
If $\E\onto\M$ is a vector bundle,
and we deal with linear connections,
then $\a:\M\to\TS\M\tn\E\tn\E^*$,
and the curvature of $\g$ can be expressed as
$$\RO[\g]\equiv -[\g,\g]=-2\,[\g_0\,,\a]-[\a,\a]~.$$
Conversely, if $\a$ is seen as the ``true'' physical field
then the choice of a gauge $\g_0$ determines the connection $\g=\g_0+\a$\,.

\remark~
If one has a local flat linear connection $\g_0$\,,
then the $\g_0$-constant local sections of $\E\onto\M$
determine a trivialization of $\E$ over any sufficiently small
open subset of $\M$.
Thus one also has $\g_0$-constant local frames.
Conversely, the assignment of a local frame determines a flat $\g_0$
by the condition that its coefficients vanish in that frame.
A gauge transformation is a section $\Ssf:\M\to\End\E$.
This transforms the family of $\g_0$-constant sections
to a new family of sections,
which determines a new flat connection\footnote{
If $\nabla[\g_0]\Ssf=0$ then $\g_0'=\g_0$
(the two families of covariantly constant sections coincide).
In that case one uses to say
that $\Ssf$ is a ``global'' gauge transformation).} 
$\g_0'=\g_0+(\nabla[\g_0]\Ssf)\pint\cev\Ssf$\,.
\smallbreak
Now the observations made in the remark concluding
\S\ref{ss:Tetrad and QED interactions}
suggest a more general,
somewhat non-standard approach to
the interactions between fermions and gauge particles.
Suppose we describe fermions as sections of a vector bundle
$$\W\equiv\W\!_{\!\Rrr}\oplus\W\!_{\!\Lll}\equiv
\bigl(\F\tn\U\bigr)\oplus\bigl(\F\tn\Ua\bigr)$$
(all tensor product and direct sums are fibered over $\M$),
where the vector bundle $\F\onto\M$
is endowed with a fibered Hermitian structure.
Then, taking into account $\U\tn\Uc=\CC\tn\H$ and $\Ul\tn\Ua=\CC\tn\H^*$,
we get natural linear inclusions
\begin{align*}
&\Wc\!_{\!\Rrr}\tn\W\!_{\!\Rrr}\equiv\Fc\tn\F\tn\H\into\Wc\tn\W~,
\\[6pt]
&\Wc\!_{\!\Lll}\tn\W\!_{\!\Lll}\equiv\Fc\tn\F\tn\H^*\into\Wc\tn\W~.
\end{align*}
Using the tetrad, the Lorentz metric of $\H$
and the Hermitian structure of $\F$,
sections of the above bundles can be seen as sections
$$\M\to\TS\M\tn\Fl\tn\F\cong\TS\M\tn\End(\F)~.$$

Let me elaborate a little further,
in order to elucidate the essential idea contained in the above considerations.
If the classical fields corresponding to certain particles are sections
of a vector bundle $\E\onto\M$,
then we also consider ``gauge particles'' interacting with them.
The classical fields corresponding to these could be, in principle, 
sections of any vector bundles whose geometric structures
yield suitable scalar-valued contractions;
given a Hermitian structure on the fibers,
$\Ec\tn\E\cong\End(\E)$ seems to be the simplest such item.
In a sense, we might see all this as roughly analogue to chemistry,
where the various kinds of atoms can bind together
according to their valences.

Now if this generic $\E$ is of the type of the above $\W$ then,
because of the structure of the spinor bundle (\S\ref{s:Electrodynamics})
and of the Hermitian structure of $\F$,
one actually recovers all the standard gauge fields
and, possibly, also further fields.\footnote{
This approach could be extended further by taking
higher tensor powers of $\W$ and of the bundles associated with it.} 
In the case of electroweak theory, in fact,
we'll (provisionally) leave out some of the occurring sectors.

Note how the Hermitian structure of $\W$ partly derives from that of $\F$,
and partly from the natural Hermitian structure of $\U\oplus\Ua$\,;
on the other hand,
like in the particular case of electrodynamics,
no Hermitian structure of $\U$ itself has to be assumed
(this would imply the choice of an observer,
see~\S\ref{ss:Two-spinors and Lorentzian geometry}).
This observation stresses an important difference between spin
and other internal symmetries,
besides the fact that only spin is directly connected to spacetime geometry
via the tetrad.

Next, continuing along this line of thought,
note that we have two further natural sub-bundles
\begin{align*}
&\Wc\!_{\!\Rrr}\tn\W\!_{\!\Lll}\equiv\Fc\tn\F\tn\Uc\tn\Ua\into\Wc\tn\W~,
\\[6pt]
&\Wc\!_{\!\Lll}\tn\W\!_{\!\Rrr}\equiv\Fc\tn\F\tn\Ul\tn\U\into\Wc\tn\W~.
\end{align*}
We see no reason why sections of the above bundles should be excluded
from the set of fields of our theory.
Actually it will turn out
that the Higgs field of electroweak theory can be described in this way.

\remark~For simplicity,
in the above preliminary discussion,
we considered the same bundle $\F$ as a tensor factor
in $\W\!_{\!\Rrr}$ and $\W\!_{\!\Lll}$\,,
but our formulation
can be easily extended to the case of a fermion bundle
of the type \hbox{$\W\equiv(\F_{\!\!\Rrr}\tn\U)\oplus(\F_{\!\!\Lll}\tn\Ua)$}\,,
where $\F_{\!\!\Rrr}\onto\M$ and $\F_{\!\!\Lll}\onto\M$
are distinct vector bundles.\smallbreak

A further important issue,
in the comparison between a classical field theory
and its pre-quantum counterpart,
is that of the \emph{scaling} of the fields.
In general the fields of a classical theory are scaled
(as we saw in the case of electrodynamics,
\S\ref{ss:Two-spinors and Einstein-Cartan-Maxwell-Dirac fields}).
Using natural units, this means that they are sections of certain
vector bundles tensorialized by powers of the unit space $\LL$\,,
needed in order to obtain, eventually, an unscaled Lagrangian density.
However scaling actually disappears in the contributions
of internal vertices and lines,
hence considering unscaled fields in the pre-quantum formulation
seems reasonable.
We may obtain the scaled version of a field by tensor product
with appropriate rational powers of the spacetime volume form $\eta$\,.

\subsection{Gauge fields and charges}
\label{ss:Gauge fields and charges}

In usual presentations,
gauge fields such as the electromagnetic potential bear
\emph{charge} factors,\footnote{In natural units, a charge is a pure number.} 
which depend on the type of the particle they interact with.
Strictly speaking this only makes sense \emph{after} a gauge has been choosen,
because multiplying a connection by an arbitrary real number is not a
geometrically well-defined operation.
The fact that the same gauge field can interact differently
with different particles is the basic reason
why one cannot simply absorb the charge into the gauge field.

Now there is, actually, a natural geometric way
of describing different charge values without choosing a gauge,
at least as long as they are all integer multiples of a basic unit.
In fact, let  $Y$ be a linear connection of a line bundle $\Q\onto\M$,
and let \mbox{$Y_{\!a}$} be its coefficient in a local frame $\bb$
(namely \hbox{$\na_a\bb=-Y_{\!a}\,\bb$});
then the coefficient of the induced connection on $\Q^p\equiv\Q\tn\cdots\tn\Q$
($p$ factors) is \hbox{$pY_{\!a}$} in the induced frame $\bb^p$.
More generally,
we can naturally accommodate for different interaction strengths
by tensorializing any vector bundle with powers of $\Q$.

A further question concerns the numeric value of the basic charge,
namely why it can't simply absorbed into the field's strength:
the answer is that this numeric value plays an essential role
in extracting the physical meaning of a theory of quantum particles.
The precise mathematical place of the basic charge
deserves a thorough discussion
(see also Weinberg~\cite{We96}, Vol.II,~\S15.1-15.2).
Gauge fields are represented as 1-forms valued into the sub-bundle
$\Lfr\subset\End\F$ of all $h$-anti-Hermitian endomorphisms
(the fibers are Lie sub-algebras).
The symmetric bilinear form\footnote{
We recall that the Killing 2-form of a Lie algebra $\Lfr$
is defined by $\KO(A,B):=\Tr(\ad_A\comp\ad_B)$\,,
where for each $A\in\Lfr$ we define $\ad_A\in\End\Lfr$ by $\ad_A(X):=A\,X$\,.
In particular, if $\Lfr$ is the Lie algebra of \emph{all} endomorphisms
of an $n$-dimensional vector space,
then $\KO(A,B)=2\,n\Tr(A\comp B)-(\Tr A)\cdot(\Tr B)$\,.
Note that the restriction of $\KO$ to a Lie subalgebra $\Lfr'\subset\Lfr$ is not,
in general, the Killing metric of $\Lfr'$.
In the case of a simple Lie algebra there is, up to a factor,
a unique invariant symmetric 2-form;
for a simple Lie algebra of endomorphisms this is
$(A,B)\mapsto\Tr(A\comp B)$\,. } 
$$\KO(X,Y):=-2\,\Tr(X\comp Y)~,\qquad X,Y\in\End\F~,$$
restricts to a positive scalar product of $\Lfr$\,;
moreover the $h$-contraction of $X,Y\in\Lfr$, defined as
$$\bang{\bar X,Y}\equiv
\bar X\Ii{\a\.}{\b\,\.}\,Y\Ii\a\b\,h_{\a\.\,\a}\,h^{\b\,\.\,\b}
=-X\Ii\b\a\,Y\Ii\a\b~,$$
can also be written as
$$\bang{\bar X,Y}=\oh\,\KO(X,Y)~.$$
Let now $\bigl(\lfr_i\bigr)$ be a $\KO$-orthonormal frame of $\Lfr$\,,
$q\in\RR\setminus\{0\}$\,,
and write $X:\M\to\TS\M\tn\Lfr$ as
$$X=q\,X_a^i\,\dx^a\tn\lfr_i~\qquad X_a^i:\M\to\RR~.$$
Then $X_a^i$ is assumed to be the physical `field strenght'.
If $X$ represents a connection of $\F$ in some chosen gauge,
then the curvature tensor of that connection has the expression
\begin{align*}
\RO[X]\equiv R\iI{ab}i\,\dx^a\we\dx^b\tn\lfr_i
&=\bigl(-q\,\de_{[a}X_{b]}^i
+q^2\,X_a^h\,X_b^j\,[\lfr_h\,,\lfr_j]^i\bigr)\,\dx^a\we\dx^b\tn\lfr_i=
\\[6pt]
&=\bigl(-q\,\de_{[a}X_{b]}^i
+q^2\,X_a^h\,X_b^j\,c\iI{hj}i\bigr)\,\dx^a\we\dx^b\tn\lfr_i~,
\end{align*}
where $c\iI{hj}i=c_{hji}\equiv\KO(\lfr_i\,,[\lfr_j\,,\lfr_k\,])$
are the `structure constants' of $\Lfr$
(the index $i$ is raised via $\KO$
and we are working with an orthonormal frame).
Next one introduces a Lagrangian density $\Lcal[X]\equiv\ell[X]\,\eta$\,,
where $\eta$ is the spacetime volume form and
$$\ell[X]:=-\frac1{2\,q^2}\,g^{ac}\,g^{bd}\,\KO_{ij}\,R\iI{ab}i\,R\iI{cd}j~,
\qquad \KO_{ij}=\d_{ij}~.$$
In this way eventually one gets `kynetic terms' of the kind
$g^{ac}\,g^{bd}\,\d_{ij}\,\de_aX_b^i\,\de_cX_d^j$
which are not affected by $q$\,,
3-factors interaction terms (of the kind
$g^{ac}\,g^{bd}\,c_{ihj}\,\de_aX_b^i\,X_c^h\,X_d^j$)
multiplied by $q$\,,
and 4-factors terms multiplied by $q^2$\,.

\section{Electroweak field theory}
\label{s:Electroweak field theory}

\subsection{The fermion bundle}
\label{ss:The fermion bundle}

The starting step for building up the geometric background
of the electroweak theory consists in the introduction of a
complex vector bundle $\W\onto\M$ whose sections
are the theory's fermion fields.
This is obtained, most naturally,
by a suitable modification of the fermion bundle
$\W\equiv\U\oplus\Ua$ of electrodynamics
(\S\ref{ss:Two-spinors and Einstein-Cartan-Maxwell-Dirac fields}).
If one chooses to mimic the usual presentations closely,
then one is led to set
$$\W\equiv\W\!_{\!\Rrr}\oplus\W\!_{\!\Lll}\equiv\U\oplus(\I\tn\Ua)~,$$
where $\I\onto\M$ is a new complex vector bundle with 2-dimensional fibers
which must be endowed with a Hermitian structure
(all tensor product and direct sums are fibered over $\M$).

However on finds at once a slight complication:
the needed charge values for gauge fields,
when seen from the point of view expressed
in~\S\ref{ss:Gauge fields and charges},
imply a precise relation between
the fibers of the bundles of ``complex volumes'' of $\U$ and $\I$\,;
namely one must have
$$\weu2\Ul\cong\weu2\I\tn\weu2\I~.$$
Equivalently, this can expressed\footnote{
If $\A$ and $\B$ are vector spaces, respectively of dimension $m$ and $n$\,,
then it is not difficult to see that there is a natural isomorphism
\hbox{$\weu{m+n}(\A\oplus\B)\cong\weu{m}\A\tn\weu{n}\B$}\,.} 
as a similar relation
$$\weu2\W\!_{\!\Rrr}\cong\weu4\W\!_{\!\Lll}$$
between the ``complex volume'' bundles of the right-handed and left-handed
fermion sectors.
I find that the most natural construction that meets the above said requirement
is obtained as follows:
first, by assuming that the two-spinor connection $\Cs$,
differently from the standard setting of electrodynamics
(\S\ref{ss:Two-spinors and Einstein-Cartan-Maxwell-Dirac fields}),
determines a curvature-free connection of $\weu2\U$\,;
second, by taking the \emph{fermion bundle} of the electroweak theory to be
$$\W\equiv\W\!_{\!\Rrr}\oplus\W\!_{\!\Lll}\equiv
\bigl(\weu2\I\tn\U\bigr)\oplus\bigl(\I\tn\Ua\bigr)~,$$
with no further assumption.

\remark~
We recover essentially the previous setting as
$$\W=\U'\oplus(\I'\tn\Uc{}'^\lin)~,$$
with $\U'\equiv\weu2\I\tn\U$ and $\I'\equiv\weu2\Ic\tn\I$.
Further suitable settings are actually possible,
the chosen one being the most natural in my opinion.
\smallbreak

We are going to formulate a field theory
in which the fields are sections valued in $\W$ and in $\Wc\tn\W$,
according to the ideas sketched
in~\S\ref{ss:Gauge fields in classical and pre-quantum theories}.
In principle there is no obstruction to including
also the gravitational field, represented by the tetrad $\Th$
and the spinor connection\footnote{
Since the induced connection of $\weu2\U$ is now assumed to be curvature-free,
the assignment of $\Cs$ is equivalent
to that of the connection $\td\G$ of $\H$.} 
 $\Cs$ of $\U$\,:
essentially, one only has to include the usual gravitational term
(\S\ref{ss:Two-spinors and Einstein-Cartan-Maxwell-Dirac fields})
in the Lagrangian density.
But for the moment we leave out this kind of extension,
and choose to work in a fixed (curved) gravitational background.
Then $\Cs$ and $\Th$ are seen as a-priori fixed structures.
A further fixed structure is the Hermitian metric of $I$,
denoted as $h:\M\to\Ia\tn\Il$ and assumed to have positive signature.

As in \S\ref{ss:Two-spinors and Lorentzian geometry}
we'll denote by $\bigl(\z_\sA\bigr)$\,, ${\scriptstyle A}=1,2$\,,
a \emph{two-spinor frame}, namely a local frame of $\U\onto\M$.
Since the induced connection of $\weu2\U$ is now assumed to be curvature-free,
we can choose the frame in such a way that
$\nabla\e\equiv\nabla(\z^1\we\z^2)=0$\,,
so that $\Cs\iIi a\sA\sA=0$
(namely $\Cs$ is purely ``gravitational'').
Note how the existence of local covariantly constant sections
$\M\to\weu2\Ul$ does \emph{not} imply that there exist a distinguished
such 2-form
(this is only unique up to a \emph{constant} phase factor).

Moreover we'll denote by $\bigl(\xi_\a\bigr)$\,, $\a=1,2$\,,
an \emph{isospin frame},
namely an $h$-orthonormal local frame of $\I\onto\M$.
We also set $\om\equiv\xi^1\we\xi^2:\M\to\weu2\Il$
(whenever no confusion arises, we distinguish dual frames
simply by upper indices).

Let $X$ be a linear connection of $\I\onto\M$.
Then its coefficients in the frame $\bigl(\xi_\a\bigr)$
can be written as
$$X\iIi a\a\b=X_a^\l\,\s\iIi a\a\b~,\qquad X_a^\l:\M\to\CC~,~~
a=1,2,3,4,~\l=0,1,2,3~.$$
One finds that the coefficients $X_a^\l$ are imaginary
if and only if $X$ fulfills the condition $\nabla[X]h=0$
(we then say that $X$ is \emph{Hermitian},
though the matrices $\bigl(X\iIi a\a\b\bigr)$
are actually \emph{anti-Hermitian}).
Namely after the choice of a gauge one views the connection
as a 1-form of $\M$ valued into the Lie algebra of $\Ug(2)$\,.

\subsection{The fields}
\label{ss:The fields}

The fermion field is a section
$$\psi\equiv\psi_{\Rrr}+\psi_{\Lll}:
\M\to\W\!_{\!\Rrr}\oplus\W\!_{\!\Lll}\equiv\W~.$$
Its coordinate expression is written as
$$\psi=\psi^\sA\,\om^{-1}\tn\z_\sA+\psi^\a_\cA\,\xi_\a\tn\bze^\cA~,$$
with $\om^{-1}=\xi_1\we\xi_2:\M\to\weu2\I$\,.

The boson fields will be introduced according to the ideas
sketched in~\S\ref{ss:Gauge fields in classical and pre-quantum theories},
with some restrictions aimed at reproducing essentially
the standard electroweak theory (and no further fields).
First we expand $\Wc\tn\W$ and reorder some tensor products,
obtaining
\begin{align*}
\Wc\tn\W~&\cong~(\Wc\!_{\!\Rrr}\tn\W\!_{\!\Rrr})~
\oplus~(\Wc\!_{\!\Lll}\tn\W\!_{\!\Lll})~
\oplus~(\Wc\!_{\!\Rrr}\tn\W\!_{\!\Lll})~
\oplus~(\Wc\!_{\!\Lll}\tn\W\!_{\!\Rrr})~=
\\[10pt]
&\cong~ (\weu2\Ic\tn\weu2\I\tn\H)~
\oplus~ (\Ic\tn\I\tn\H^*)~\oplus~\\[6pt]
&\qquad~\oplus~ (\weu2\Ic\tn\I\tn\End\Uc)~
\oplus~ (\weu2\I\tn\Ic\tn\End\U)~.
\end{align*}

We first focus our attention to the two last bundles,
which are mutually conjugate.
We'll consider sections having the special form
$$\phi\tn\Id{\Uc}:\M\to\weu2\Ic\tn\I\tn\End\Uc~~
\text{where}
~~\phi:\M\to\weu2\Ic\tn\I$$
(here
$\Id{\Uc}:\M\to\Uc\tn\Ua\cong\End\Uc$ denotes identity of $\Uc\onto\M$).
Then also
$$\bar\phi\tn\Id{\U}:\M\to\weu2\I\tn\Ic\tn\End\U~.$$

A section $\phi$ as above will be called a \emph{Higgs field}.
We could consider more general fields in this sector,
by dropping the restriction of proportionality to the identity,
but for the moment we won't broaden our investigation too much.

The gauge fields are represented by a section
$$W:\M\to\Ic\tn\I\tn\H^*\cong\Wc\!_{\!\Lll}\tn\W\!_{\!\Lll}~.$$
By contraction with the Hermitian metric $h$ of $\I$ this also yields
a section
$$\widehat W:\M\to\weu2\Ic\tn\weu2\I\tn\H^*~,$$
which, using the natural Lorentz metric of $H$, can also be seen as a section
$$\M\to\Wc\!_{\!\Rrr}\tn\W\!_{\!\Rrr}\cong\weu2\Ic\tn\weu2\I\tn\H~.$$
More precisely, we indicate as $\hat h:\M\to\weu2\Ia\tn\weu2\Il$
the Hermitian metric induced on $\weu2\I$
(then $\hat h$ is a positive object), and set
$$\widehat W:=\hat h{}^\#\tn (h\pint W):\M\to\weu2\Ic\tn\weu2\I\tn\H^*~.$$
In order to express the above field in component notation
we consider the frame $\bigl(\io_\l\bigr)$
of $\Ic\tn\I$ defined by
$$\io_\l:=\s_\l^{\a\a\.}\,\bar\xi_{\a\.}\tn\xi_\a~,\quad\l=0,1,2,3$$
(then $\bang{\io_\l\,,\io_\m}=2\,\eta_{\l\m}$),
and write
$$W=W^\m_\l\,\io_\m\tn\t^\l \qRq \widehat W=2\,W_\l^0\,\hat h{}^\#\tn\t^\l~.$$

If the field $W$ is known,
then the choice of an isospin gauge determines a isospin connection
with components $X\iIi a\a\b=\iO\,q\,W_\l^\m\,\s\iIi\m\a\b$\,;
the (unique) component of the induced connection $\widehat X$ of $\weu2\I$
is $\widehat X_a=2\,\iO\,q\,W_\l^0$\,.

\subsection{Symmetry breaking}
\label{ss:Symmetry breaking}

At this point we introduce the symmetry breaking
associated with the Higgs field according
to the standard view (more or less):
one assumes that there is one special section
$\phvac:\M\to\weu2\Ic\tn\I$,
seen as the ``vacuum expectation value'' of $\phi$\,,
supposedly arising as a minimum of the ``Higgs potential''
$$V[\phi]:=\l\,(2\,\m^2\,\bang{\bar\phi,\phi}-\bang{\bar\phi,\phi}^2)~,$$
where $\l,\m\in\RR^+$.
We can choose the $h$-orthonormal isospin frame $\bigl(\xi_\a\bigr)$
in such a way that $\phvac=\m\,\xi_2$\,;
thus $\xi_2$ is determined by $\phvac$\,,
while $\xi_1$ is determined up to a phase factor.
Next we observe that there is a unique $h$-preserving endomorphism
$\Ssf_\phi:\M\to\End\I$ such that
$\phi=\Ssf_\phi(\|\phi\|\,\xi_2)$\,,
with $\|\phi\|\equiv\bang{\bar\phi,\phi}{}^{1/2}$\,;
its matrix $(\Ssf\Ii\a\b)$ in any $h$-orthonormal frame
is valued into $\SU(2)$\,.
Moreover we write $f\equiv\|\phi\|\,{-}\,\m$
and represent $\phi$ as the couple of fields
$$(f,(\Ssf_\phi)):\M\to\RR\times\SU(2)~.$$

In the classical description, $\Ssf_\phi$ is usually ``absorbed'' into
the isospin connection through the following argument.
Let $\bigl(\xi'_\a\bigr)\equiv\bigl(\Ssf_\phi(\xi_\a)\bigr)$
be the ``rotated'' frame; then
$\phi=\|\phi\|\,\xi_2'\equiv(\m\,{+}\,f)\,\xi_2'$\,.
We also express the left-handed fermion field in the new frame,
namely we write $\psi_{\Lll}=\psi'{}^\a_\cA\,\xi'_\a\tn\bze^\cA$
with $\psi'{}^\a_\cA\equiv\cev\Ssf\Ii\a\b\,\psi^\b_\cA$\,.
At the same time we consider the \emph{new} connection $X'$
whose components $X'{}\iIi a\a\b$ in the frame $\bigl(\xi'_\a\bigr)$
are the same\footnote{
Namely $X'$ is characterized by
$\na_a[X'](\Ssf_\phi\s)=\Ssf(\na_a[X]\s)$ for all sections $\s:\M\to\I$.
The components of $X'$ in the frame $\bigl(\xi_\a\bigr)$ are
$\bigl(\Ssf\,X_a\,\cev\Ssf+(\de_a\Ssf)\,\cev\Ssf\bigr)\Ii\a\b$\,.} 
as the components of $X$ in the frame $\bigl(\xi_\a\bigr)$\,.
The two connections have the same curvature tensor,
but the gauge freedom of $X'$ is reduced;
now, in fact, we only obtain a classically equivalent theory
by an $\Ug(1)$ gauge transformation affecting $\xi'_1$\,,
while $\xi'_2$ is fixed.
Eventually, one drops all the primes and says that the 3 real degrees of freedom,
eliminated from $\phi$ through the gauge transformation $\Ssf_\phi$\,,
have been ``eaten up'' by the connection.

The vacuum value of the Higgs field selects a subbundle
$\II\subset\I\onto\M$ whose fibers are the positive spaces
generated by that value.
Thus symmetry breaking determines a decomposition of $\I$
into $h$-orthogonal subbundles
$$\I=\I_1\oplus\I_2=\I_1\oplus(\CC\tn\II)~,$$
with $\xi_1:\M\to\I_1$ and $\xi_2:\M\to\II$\,.

Let $\td h\equiv\bh\tn h$ denote the Hermitian metric of $\Ic\tn\I$
determined by $h$\,,
and observe that the inverse of $h$ itself is a section
$h^\#:\M\to\Ic\tn\I$\,.
Hence $\Ic\tn\I$ can be decomposed as the direct sum
$$\Ic\tn\I=(\Ic\tn\I)_h~\oplus~(\Ic\tn\I)_h^{{}^\bot}~,$$
of the sub-bundles constituted of all elements proportional to $h^\#$
and of all elements orthogonal to $h^\#$, respectively.
Symmetry breaking determines a further decomposition of $\Ic\tn\I$
into four mutually $\td h$-orthogonal subbundles
with complex 1-dimensional fibers:
$$\Ic\tn\I=(\,\Ic_1\tn\I_1)~\oplus~(\,\Ic_2\tn\I_2)~
\oplus~(\,\Ic_1\tn\I_2)~\oplus~(\,\Ic_2\tn\I_1)~.$$

In order to describe the gauge fields of e.w.\ theory
we are also interested in \emph{real} subbundles of $\Ic\tn\I$
with 1-dimensional fibers.
We have, of course, the Hermitian subbundle $\E_0\subset(\Ic\tn\I)_h$\,,
that is the real bundle generated by $h^\#$,
and the Hermitian subbundles
$\E_1\subset\Ic_1\tn\I_1$ and $\E_2\subset\Ic_2\tn\I_2$\,.
Then
$$\bar\xi_1\tn\xi_1=\oh\,(\io_1+\io_3)~,\quad
\bar\xi_2\tn\xi_2=\oh\,(\io_1-\io_3)~,$$
are distinguished frames of $\E_1$ and $\E_2$\,,
respectively.\footnote{Here also expressed in terms of the above introduced
(\S\ref{ss:The fields}) frame
$\bigl(\io_\l\bigr)\equiv\bigl(\s_\l^{\a\a\.}\,\bar\xi_{\a\.}\tn\xi_\a\bigr)$\,.} 

We'll need a further real subbundle
$\E'\subset\E_1\oplus\E_2$\,.
This is not completely determined by the geometric structures
assumed up to now,
but needs a new `ingredient': the \emph{Weinberg angle} $\thW\in(0,\pi/2)$\,.
Now $\E'$ is defined as the subbundle of $\E_1\oplus\E_2$ generated by
$$\io':= -\oh\,\bar\xi_1\tn\xi_1+\oh\,\cos(2\,\thW)\,\bar\xi_2\tn\xi_2
\equiv -\oh\,\bigl[\sin^2(\thW)\,\io_0+\cos^2(\thW)\,\io_3 \bigr]~.$$

The subbundle of $\Ic\tn\I$ composed of all elements
orthogonal to $\E_1$ and $\E_2$ is
$$\E^+\oplus\E^-\equiv(\,\Ic_1\tn\I_2)\oplus(\,\Ic_2\tn\I_1)~.$$
These subbundles $\E^\pm$ have complex 1-dimensional fibers,
and we do not try to select a real subbundle.
We observe that the map $w\to w^\dag$ (\S\ref{ss:Hermitian spaces})
determines an anti-isomorphism \hbox{$\E^+\leftrightarrow\E^-$}.

Eventually, the gauge fields of electroweak theory are the real fields
$$A:\M\to\H^*\tn\E_2~,\quad Z:\M\to\H^*\tn\E'~,$$
and the complex fields
$$W^+:\M\to\H^*\tn\E^+~,\quad W^-:\M\to\H^*\tn\E^-~.$$
However $W^-$ is assumed to be the Hermitian adjoint of $W^+$
so that we actually have just one independent complex field.
The coordinate expressions of the gauge fields are
\begin{align*}
& A=A_\l\,\t^\l\tn\bar\xi_2\tn\xi_2~,
\\[6pt]
& Z=Z_\l\,\t^\l\tn\io'\equiv
\oh\,Z_\l\,\t^\l\tn
\bigl(-\bar\xi_1\tn\xi_1+\cos(2\,\thW)\,\bar\xi_2\tn\xi_2\bigr)~,
\\[6pt]
& W^+=W^+_\l\,\t^\l\tn\bar\xi_1\tn\xi_2\equiv
\oh\,W^+_\l\,\t^\l\tn\bigl(\io_1+\iO\,\io_2 \bigr)~,
\\[6pt]
& W^-=W^-_\l\,\t^\l\tn\bar\xi_2\tn\xi_1\equiv
\oh\,W^-_\l\,\t^\l\tn\bigl(\io_1-\iO\,\io_2 \bigr)~,
\end{align*}
with
$$W^-_\l=\bar W_\l^+:\M\to\CC~.$$

Thus the set of gauge fields is composed essentially
by two real fields and one complex field.

\remark~
One could introduce the gauge-field target bundle
$\E\equiv\E_2\oplus\E'\oplus\E^+$
as fully unrelated to the spin and isospin bundles,
and describe the interactions via further geometric assumptions
(see also~\cite{De02}).
\smallbreak

\subsection{Electroweak geometry and the dilaton}
\label{ss:Electroweak geometry and the dilaton}

As we saw in
\S\ref{ss:Two-spinors and Einstein-Cartan-Maxwell-Dirac fields},
the ``breaking of local conformal invariance'' occurring in real physics
actually amounts to the choice of a curvature-free connection $G$
of the bundle $\LL\onto\M$ of length units.
We also saw how a natural candidate for $\LL$ arises
in the context of electrodynamics.
In the extended context of electroweak geometry
one could associate $\LL$ with different constructions,\footnote{
For example, one could see $\LL$
as the positive Hermitian subspace of $\weu2\Ic\tn\weu2\I$.} 
but that is not an essential point since
any space can be tensorialized by arbitrary powers of $\LL$\,.
The point, instead, is that the various terms in the Lagrangian density
must be ``padded'' with powers of $\LL$\,,
so that all have the same ``conformal weight''.
This is unavoidable because some of the fields,
notably the metric and the induced volume form,
are inherently scaled.\footnote{
One can set up a more general ``running constants'' formalism
by assuming bundles $\MM\,,\LL\,,\TT\onto\M$
and letting sections 
$c:\M\to\LL\tn\TT^{-1}$\,, $\h:\M\to\MM\tn\LL^2\tn\TT^{-1}$
not to be fixed.} 

In the Einstein-Cartan-Maxwell-Dirac field theory,
the curvature-free connection $G$ of $\LL\onto\M$
must be just assumed,
while in the context of electroweak geometry one is intrigued to look
for possible mechanisms determining the breaking of local conformal invariance
in relation to the Higgs field and the breaking of isospin symmetry
(also encouraged by the evasive nature of the Higgs).
Recent proposals suggest seeing the $h$-norm
of the Higgs field as a conformal factor
for the spacetime metric~\cite{Fa,FKD,RySh},
or the Hermitian metric of the isospin bundle itself as an independent field
rather than a fixed structure~\cite{Ta}.
Here I wish to frame the question in the context
of the above said point of view of determining
a curvature-free connection $G$\,.
Then I'll argue, somewhat differently from other proposals,
that any convincing solution of the said question would require
a substantial extension of the electroweak theory.

Let's start from the Lagrangian density of the electroweak theory
in a ``classical'' field context;
as usual we write it as the sum
$$\Lcal=\Lcal_\psi+\Lcal_\phi+\Lcal_X+\Lcal_{\mathrm{int}}~,$$
with
\begin{align*}
\Lcal_\psi&=\isq\sTh\Bigl(
\nabla\psi_\Rrr\tn\bar\psi_\Rrr-\psi_\Rrr\tn\nabla\bar\psi_\Rrr
+\td g^\#(\bang{\nabla\psi_\Lll\tn\bar\psi_\Lll}
-\Bang{\psi_\Lll\tn\nabla\bar\psi_\Lll})\Bigr)~,
\\[6pt]
\Lcal_\phi&=\bigl(\bang{g^\#\tn\nabla\bar\phi\tn\nabla\phi}
+\l(2\,m^2\,\bang{\bar\phi\tn\phi}-\bang{\bar\phi\tn\phi}^2)\bigr)\,\eta~,
\quad m\in\LL^{-1},~\l\in\RR^{+}~,
\\[6pt]
\Lcal_X&=-\bang{g^\#\tn g^\#\tn R[\bar X]\tn R[X]} \eta~,
\\[6pt]
\Lcal_{\mathrm{int}}&=
-(\bang{\bar\psi_\Lll\tn\phi\tn\psi_\Rrr}
+\bang{\bar\psi_\Rrr\tn\bar\phi\tn\psi_\Lll})\,\eta~.
\end{align*}
Here, $R[X]$ denotes the curvature tensor of the connection $X$ of $\I\onto\M$,
the angular bracket denotes $h$-contraction
and $\sTh$
is the tetrad-related operation introduced
in~\S\ref{ss:Two-spinors and Einstein-Cartan-Maxwell-Dirac fields}.
In coordinates, using an $h$-orthonormal frame $\bigl(\xi_\a\bigr)$ of $\I$,
we get
$\Lcal=(\ell_\psi+\ell_\phi+\ell_X+\ell_{\mathrm{int}})\,\dO^4\xx$\,,
where
\begin{align*}
\ell_\psi&=
\isq\,\sTh^a_{\AAd}\,\Bigl(\na_a\psi^\sA\,\bps^\cA-\psi^\sA\,\na_a\bps^\cA
+\e^{\sA\sB}\be^{\cA\cB}\,
(\bps_{\sB\a}\,\na_a\psi_\cB^\a-\na_a\bps_{\sB\a}\,\psi_\cB^\a\,)
\Bigr)~,
\displaybreak[2]\\[6pt]
\ell_\phi&=\Bigl( g^{ab}\,\na_a\bar\phi_\a\,\na_b\phi^\a
+2\,\l\,m^2\,(\bar\phi_\a\,\phi^\a)-\l\,(\bar\phi_\a\,\phi^\a)^2 \Bigr)\,
\det\Th~,
\displaybreak[2]\\[6pt]
\ell_X&=-g^{ac}\,g^{bd}\,\bar R_{ab\,\a\a\.}\,R\iI{cd}{\a\a\.}\,\det\Th~,
\qquad \ell_{\mathrm{int}}=-\Bigl(
\bps_{\sA\a}\,\phi^\a\,\psi^\sA+\bps^\cA\,\bar\phi_\a\,\psi_\cA^\a
\Bigr)\,\det\Th
\end{align*}
(shorthands
$\bps_{\sB\a}\equiv h_{\a\.\,\a}\,\bps_\sB^{\a\.}$\,,
$\bar\phi_\a\equiv h_{\a\.\,\a}\,\bar\phi^{\a\.}$
and the like were used).

Now one must check that all the terms in $\Lcal$ bear the same
conformal weight.
The most natural way to achieve this is by assuming
that all terms are conformally invariant, namely non-scaled
(this is the standard assumption anyway).
Taking into account the scaling (or ``conformal weight'')
of the involved objects
we find that the Fermion field and the Higgs field must be respectively
$\LL^{-3/2}$-scaled and $\LL^{-1}$-scaled, namely
\begin{align*}
&\psi\equiv\psi_\Rrr+\psi_\Lll:\M\to\LL^{-3/2}\tn\W
\equiv\LL^{-3/2}\tn\bigl(\W\!_{\!\Rrr}\oplus\W\!_{\!\Lll}\bigr)~,
\\[6pt]
&\phi:\M\to\LL^{-1}\tn\weu2\Ic\tn\I~.
\end{align*}
Hence also $m\in\LL^{-1}$, and
$$\|\phi\|^2\equiv\bang{\bar\phi\tn\phi}:\M\to\LL^{-2}~.$$

All the above setting makes sense even if conformal symmetry is not broken,
with the only addendum that any scaled coupling factors
can't be seen as constants;
in particular $m:\M\to\LL^{-1}$.
Any given section $\r:\M\to\LL^r$, $r\in\ZZ\setminus\{0\}$\,,
determines a curvature-free connection of $\LL\onto\M$
by the condition $\nabla\r=0$\,.
It's then clear that a ``vacuum value'' of $\phi$\,,
namely a special section $\phvac:\M\to\LL^{-1}\tn\weu2\Ic\tn\I$,
also determines such a connection
(fulfilling $\nabla\|\phvac\|^2=0$).
This observation, however, does not settle our question,
since assigning the Higgs potential
$\l(2\,m^2\,\bang{\bar\phi\tn\phi}-\bang{\bar\phi\tn\phi}^2)$
also means fixing the section $m$\,;
namely, the connection $G$ is already determined
by the condition $\nabla[G]m=0$\,.
Now rather than shifting between essentially equivalent choices,
which is not so interesting,
we'd like to find an independent mechanism.

Also note that, more generally, if $\phi$ is indeed scaled then we can't
write a polynomial expression in $\|\phi\|$ and/or $\|\phi\|^{-1}$,
having any positive root, without fixing some scaled factor.
On the other hand, if we let $m$ (say) be an independent field,
variation of the Lagrangian with respect to it
yields a relation between $m$ and $\phi$ but no condition on $G$.

So we see that the standard setting of the electroweak theory is not sufficient
to give the question of conformal symmetry breaking a final answer.
Instead we should extend the theory in some way.
The most obvious extension would be allowing $G$ as a dynamical field
and adding to the Lagrangian the non-scaled term
$$\bang{g^\#\tn g^\#,\dO G\tn\dO G}\,\eta\equiv
g^{ac}\,g^{bd}\,\de_aG_b\,\de_cG_d\,\det\Th\,\dO^4\xx~.$$
Now $G_a$ also appears in the covariant derivatives
of the various scaled fields;
these act as sources for $G$\,, and we cannot expect $\dO G=0$\,.
Moreover, since the spacetime metric is scaled,
it seems that gravitation must be fully included in the dynamical problem;
with regard to this aspect,
let me say that I'm not convinced by proposals which involve
expressing the metric as a fixed background metric
multiplied by some further conformal factor
(at least, that would be against
Ockam's principle--entities shouldn't be multiplied without necessity).
For similar reasons,
if one is considering the generalization
of viewing the Hermitian metric $h$ of $\I$ as an independent field
(e.g.\ see Talmadge's proposal~\cite{Ta}),
I'd rather not take the new field as a variation of some fixed background.

\vfill\newpage

\end{document}